\documentclass[amsmath,amssymb,superscriptaddress,aps,a4paper,prl,twocolumn]{revtex4-1}

\usepackage{graphicx} 
\usepackage{amsmath}
\usepackage{amssymb}
\usepackage{amsfonts}
\usepackage{lmodern}
\usepackage{textcomp}
\usepackage[T1]{fontenc}
\usepackage{gensymb} 
\usepackage[seperr, load={}]{siunitx}

\begin{document}

\title{Probing the local temperature of a 2DEG microdomain with a quantum dot: measurement of 
electron-phonon interaction}

\author{S. Gasparinetti}
	\email{simone@boojum.hut.fi}
	\affiliation{NEST, Istituto Nanoscienze-CNR and Scuola Normale Superiore, Piazza S. Silvestro 12, I-56127 Pisa, Italy}
	\affiliation{Low Temperature Laboratory, Aalto University, P.O. Box 15100, FI-00076 Aalto, Finland}
\author{F. Deon}
	\affiliation{NEST, Istituto Nanoscienze-CNR and Scuola Normale Superiore, Piazza S. Silvestro 12, I-56127 Pisa, Italy}
\author{G. Biasiol} 
		\affiliation{CNR-IOM, Laboratorio TASC, Area Science Park, I-34149 Trieste, Italy}
\author{L. Sorba}
	\affiliation{NEST, Istituto Nanoscienze-CNR and Scuola Normale Superiore, Piazza S. Silvestro 12, I-56127 Pisa, Italy}
\author{F. Beltram}
\affiliation{NEST, Istituto Nanoscienze-CNR and Scuola Normale Superiore, Piazza S. Silvestro 12, I-56127 Pisa, Italy}
\author{F. Giazotto}
\email{giazotto@sns.it}
	\affiliation{NEST, Istituto Nanoscienze-CNR and Scuola Normale Superiore, Piazza S. Silvestro 12, I-56127 Pisa, Italy} 

\begin{abstract}
We demonstrate local detection of the electron temperature in a two-dimensional microdomain using a quantum dot.
Our method relies on the observation that a temperature bias across the dot changes the functional form of Coulomb-blockade peaks. We apply our results to the investigation of electron-energy relaxation at subkelvin temperatures, find that the energy flux from electrons into phonons is proportional to the fifth power of temperature and give a measurement of the coupling constant.
\end{abstract}

\maketitle

The role of temperature and the investigation of heat transport in solid-state nanoscale systems are currently under the spotlight \cite{Giazotto2006}. Electronic thermometers \cite{Pekola1994,Appleyard1998,Hoffmann2009}, refrigerators \cite{Prance2009, Tirelli2008, Timofeev2009a, Kafanov2009} and heat transistors \cite{Saira2007} based on metallic and semiconductor nanostructures were recently demonstrated. Such devices are typically operated at subkelvin temperatures, where electron-phonon interaction is weak and, as a result, the relevant electronic temperature may significantly differ from that of the lattice.
In this context, the ability to perform local measurements of the electronic temperature becomes highly desirable.
While in metallic systems superconducting tunnel junctions are routinely employed to this end, for the case of two-dimensional electron gases (2DEGs) or semiconductor nanowires it was shown that quantum dots (QDs) can serve the purpose \cite{Beenakker1991,Kouwenhoven1997}, due to the fact that at thermal equilibrium the linewidth of zero-bias Coulomb blockade (CB) peaks in the weak-coupling regime is directly related to the electronic temperature of the leads.

In this Letter we propose a method for the detection of local temperature in a 2DEG based on the analysis of the CB-peak lineshape in the presence of a temperature bias, and apply it to the investigation of energy-relaxation mechanisms in a micrometer-sized electronic domain. The domain is defined electrostatically and connected to the surrounding 2DEG regions through a QD and three quantum point contacts (QPCs). As a known heating power is delivered to the domain, conductance across the QD is probed and the steady-state temperature detected. The dependence of the measured electron temperature on the heating power is studied for different values of the QPC resistances: as the coupling to the surrounding 2DEG regions is reduced, we observe the crossover from a regime where excess heat is carried away by hot quasiparticles tunneling through the QPCs to one where power exchange with lattice phonons dominates. Our measurements show that the latter mechanism follows the $T^5$ power law expected \cite{Ma1991} for the screened electron-acoustic phonon piezoelectric interaction and yielded a measurement of the coupling constant.

\begin{figure}\label{Fig1}
\includegraphics{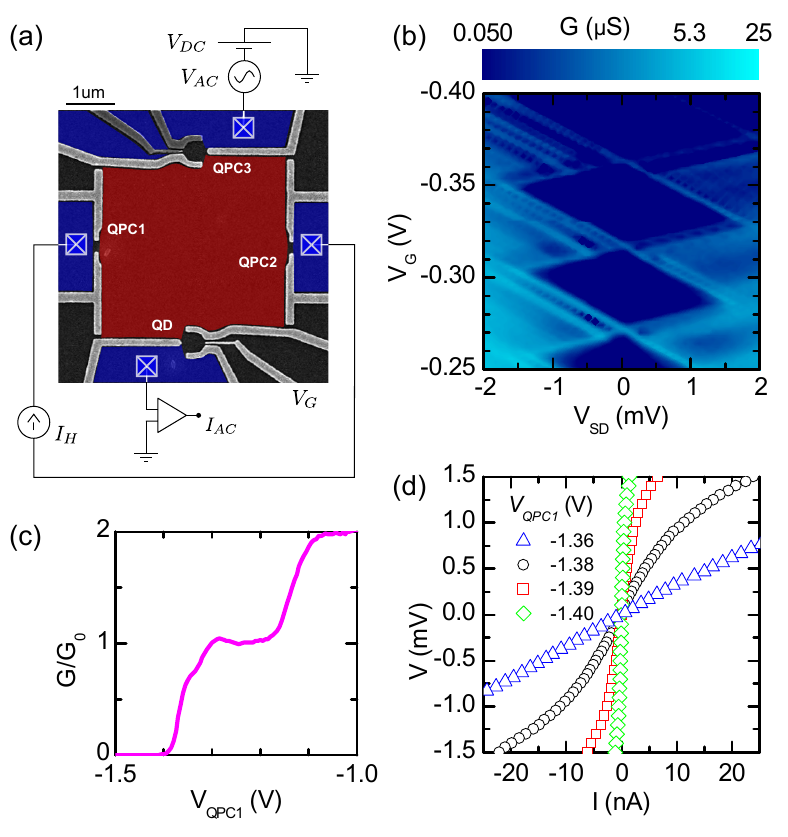}
\caption{(color online). (a) False-color scanning electron micrograph of the device and scheme of the measurement setup. Schottky gates (light gray) define a \SI{16}{\micro m^2} sized central electronic domain connected to outer regions of the 2DEG through a QD (bottom) and three QPCs (left, right, top). Crossed squares indicate Ohmic contacts to the 2DEG. (b-d) Characterization measurements: finite-bias stability plot of the QD (b), zero-bias conductance $G$ of QPC1 versus gate voltage $V_{QPC1}$ (c), and current-voltage characteristics of QPC1 for different values of the gate voltage $V_{QPC1}$ close to pinchoff (d). $G_0$ is the conductance quantum.}
\end{figure}

Figure 1(a) shows a scanning electron micrograph of the device. Aluminum surface Schottky gates were patterned on the GaAs/AlGaAs heterostructure by electron beam lithography, thermal evaporation and liftoff. The 2DEG is characterized by a density $n_s=\SI{2.26e11}{cm^{-2}}$ and a mobility $\mu=\SI{3.31e6}{cm^2/Vs}$.
When a negative voltage is applied to surface gates, a central domain of area $A_D=\,$\SI{16}{\mu m^2} is defined by lateral confinement together with a QD and three QPCs.
The measurement configuration is schematically shown in the same Fig.~1(a).
Experiments were carried out in a $^3$He cryostat down to 250 mK.
The domain was heated by driving a dc current $I_H$ with a floating source through the QPC1-domain-QPC2 circuit while the zero-bias differential conductance was simultaneously measured across the series QPC3-domain-QD
\footnote{We used \SI{15}{\micro eV} excitation voltage at low frequency (\SI{7.8}{Hz}).
QPC3 was set to be much less resistive than the QD, so that its voltage drop can be neglected.}.

The single-particle energy-level spacing associated with the lateral dimensions of the domain [$\pi^2\hbar^2/(2m^*A_D)\approx$ \SI{400}{neV}] is much smaller than the thermal energy ($k_B T\approx \SI{21}{\micro eV}$ at \SI{250}{mK}, where $k_B$ is the Boltzmann constant and $m^*$ is the effective mass), so that electrons in the domain can be treated as a Fermi gas.
The charging energy of the domain was also found to have negligible impact on the low-bias measurements at the temperatures here of interest.
Figures 1(b-d) show transport measurements performed on the QD and QPC1 with all other gates set to ground.
The QD geometry was chosen in order to maximize its charging energy $E_C$ and its single-particle energy spacing $\delta E$. In the diamonds with lowest occupation number we measured $E_C=\SI{1.5}{meV}$ and $\delta E\approx\SI{0.4}{meV}$ [see Fig.~1(b)]. 
All QPCs show clear zero-bias conductance steps, as shown in Fig.~1(c) for the case of QPC1. 
Fig.~1(d) shows QPC1 current-voltage characteristics at different gate voltages: as the gate voltage approaches pinchoff curves become increasingly non-linear in the given current interval.
\begin{figure}\label{fig2}
\includegraphics{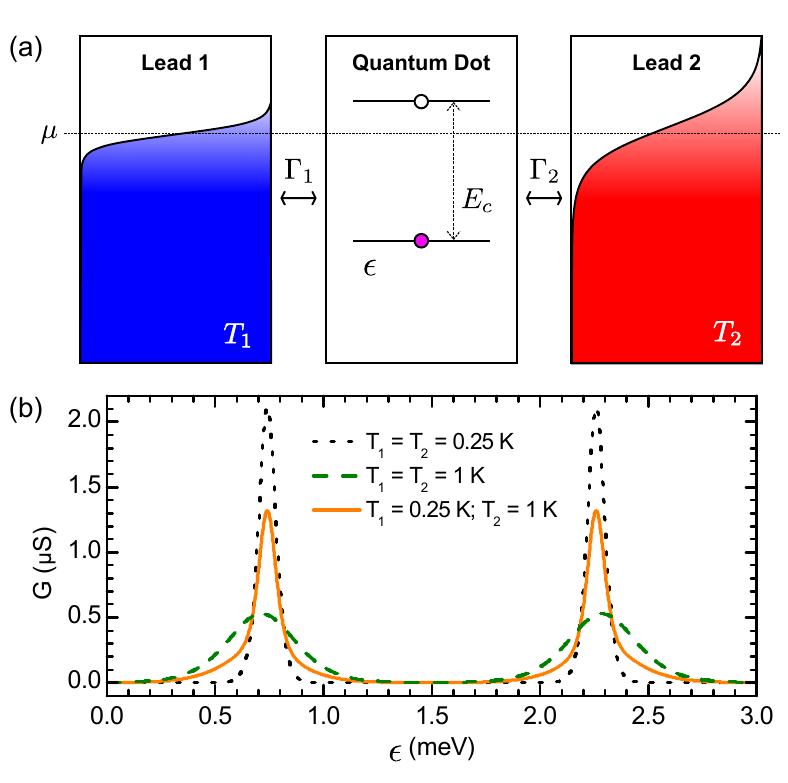}
\caption{(color online). (a) Energy-level diagram for transport through a temperature-biased QD (all quantities are defined in the text). (b) Calculated zero-bias conductance $G(\epsilon)$ for different temperatures of the leads: $T_1=T_2=\SI{250}{mK}$ (dotted line), $T_1=T_2=\SI{1}{K}$ (dashed line), $T_1=\SI{250}{mK}$ and $T_2=\SI{1}{K}$ (full line). }
\end{figure} 

Based on the framework of Ref.~\cite{Beenakker1991}, we developed a model describing the CB-peak lineshape when a finite temperature bias is applied across the QD. We assumed for the QD a spin-degenerate single-particle energy spectrum and an orbital level spacing exceeding thermal energy. As a consequence, only one orbital level contributes to conduction for a given peak and we can model the QD as an Anderson impurity \cite{Anderson1961,Meir1991} with energy level $\epsilon$ and on-site interaction $E_C$. The dot is tunnel-coupled to ideal leads at chemical potential $\mu_1$ ($\mu_2$) and temperature $T_1$ ($T_2$). Coupling is assumed to be weak, so that transport is dominated by sequential tunneling and finite-lifetime broadening effects can be neglected. The energy diagram of the modeled system is sketched in Fig.~2(a).
In the limit of large charging energy ($E_C \gg k_B T_{1,2}$) and close to the resonance $\mu_1 = \mu_2 = \epsilon$,
the current is given by
\begin{equation}\label{eq:curr_deg}
 I(\epsilon)=-2e \Gamma_1 \Gamma_2 \frac{ f_1(\epsilon)- f_2(\epsilon)}
{\Gamma_1 \left[1+ f_1(\epsilon) \right] + \Gamma_2 \left[1+ f_2(\epsilon) \right]} ,
\end{equation}
where $f_i(\epsilon)=\left[1+e^{{(\epsilon-\mu_i)/k_B T_i}}\right]^{-1}$ is the Fermi-Dirac distribution function of the $i^{\mathrm{th}}$ lead, $\Gamma_i$ is the tunneling rate between the dot and the $i^{\mathrm{th}}$ lead and $e$ is the electron charge
\footnote{A similar expression holds when $\mu_1=\mu _2=\epsilon +E_C$.}.
Fig.~2(b) shows the calculated $G$ versus $\epsilon$ in the degenerate case for some representative values of $T_1$ and $T_2$, assuming $E_C=1.5$ meV. When $T_1$ = $T_2$ temperature variations rescale peaks without affecting their shape (black, green lines). By contrast, when $T_1 \neq T_2$ the line-shape is distorted: peak tails are primarily determined by the higher temperature while the peak body by the lower one (orange line).

\begin{figure}\label{fig3}
\includegraphics[]{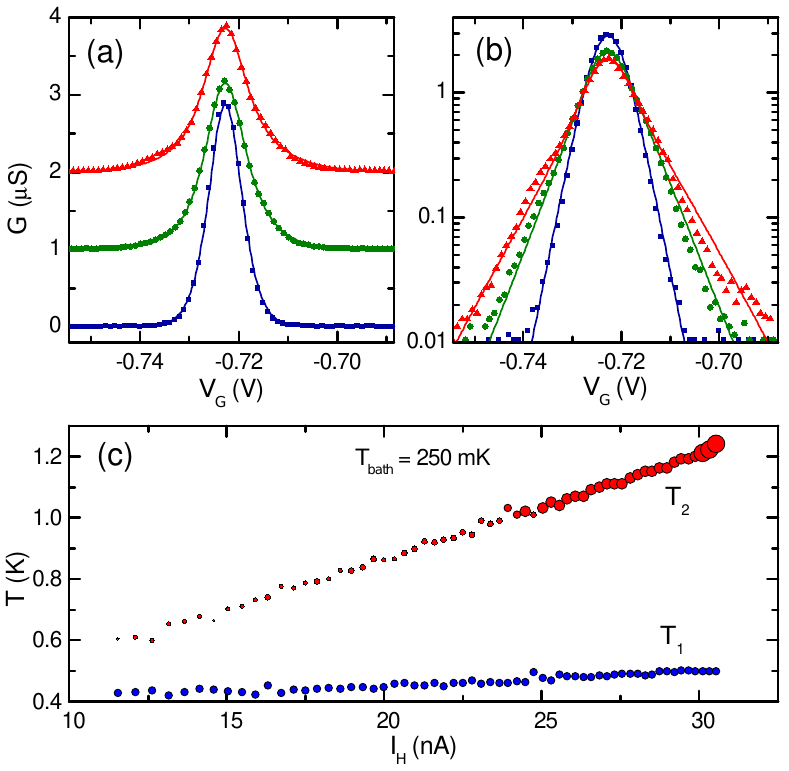}
\caption{(color online). (a,b) Conductance $G$ versus gate voltage $V_G$ for increasing values of the heating current $I_H$: \SI{11.6}{nA} (squares), \SI{20}{nA} (circles), \SI{28.6}{nA} (triangles). The corresponding full lines are best-fits of the derivative of Eq.~\ref{eq:curr_deg} to each dataset. The hotter and colder temperature thereby estimated are: $T_1=\SI{426}{mK}, T_2=\SI{604}{mK}$, $T_1=\SI{446}{mK}, T_2=\SI{861}{mK}$ and $T_1=\SI{485}{mK}, T_2=\SI{1.15}{K}$,
respectively. The data are plotted on both linear (a) and semi-logarithmic scale (b); in (a) traces have been offset vertically for clarity.
(b) Extracted temperatures $T_1$ and $T_2$ as a function of $I_H$. The hotter $T_2$ pertains to the heated microdomain, the colder $T_1$ to the external portion of the 2DEG.}
\end{figure}

The presence of this qualitatively different behavior is very useful for a comparison with experimental results. In doing so, we shall assume that the fast electron-electron scattering rate \cite{Giuliani1982} does establish a quasi-equilibrium regime for the confined electron system, so that the microdomain can play the role of the hotter lead.
Figures 3(a,b) show the measured zero-bias conductance $G$ versus gate voltage $V_G$ at T$_{bath}=250$ mK. The different curves correspond to increasing heating currents $I_H$ driven through the QPC1-QPC2 circuit. Here, the device is tuned so that $R_{QPC1,2,3}\simeq \SI{7.4}{k\ohm}$, while the conductance of the QD is of the order of a few \SI{}{\micro S}.
These experimental characteristics closely resemble the calculated curves of Fig.~2(b). In order to perform a quantitative comparison, the derivative of Eq.~\ref{eq:curr_deg} was fitted to the experimental points. The QD barriers were tuned to be symmetric, therefore we put $\Gamma_1=\Gamma_2$ in the least-squares fit. Conversion of the gate voltage $V_G$ into the dot energy $\epsilon$ was provided by the finite-bias stability plot \cite{Kouwenhoven1997}. Finally, after normalizing the curve to the position and value of its maximum, the two lead temperatures are the only free parameters.
Following this procedure we can associate two temperatures ($T_1$ and $T_2$, with $T_2 > T_1$) to each measured conductance characteristic. Figure 3(c) shows the extracted values of these temperatures as a function of the injection current $I_H$.
$T_2$ (in red) steadily increases with $I_H$, while $T_1$ (in blue) exhibits only a small change, its value remaining close to the one measured in an open configuration where only the gates defining the QD are polarized. We hence assign $T_2(I_H)$ to the central domain and $T_1(I_H)$ to the adjacent 2DEG region.
There is a $\sim 150$ mK difference between the cryostat base temperature and the electronic temperature in the absence of injection that we ascribe to power leaking through cryostat lines. The additional increase of $T_1$ with $I_H$ stems from the global device heating produced by large injection currents.
The present procedure allowed us to measure local temperature differences from about 150 mK up to about 1 K. When $T_2-T_1\lesssim\SI{150}{mK}$, the fitting algorithm fails due to the proximity of two symmetrical local minima in the parameter space.

\begin{figure}\label{fig4}
\includegraphics[]{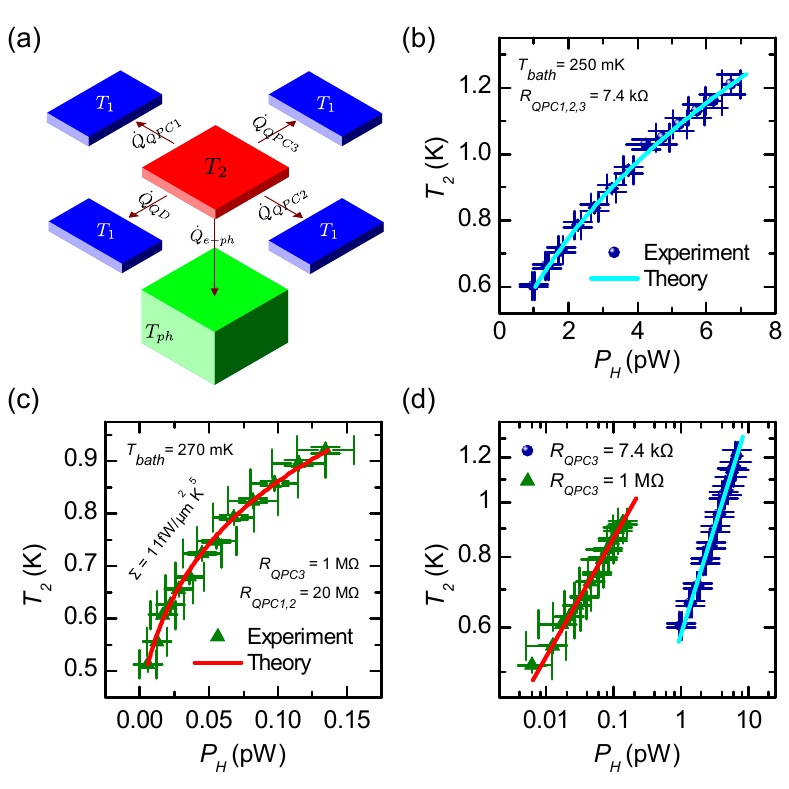}
\caption{(color online) (a) Scheme of the main contributions to the steady-state power balance.
(b-c) Temperature $T_2$ versus injected power $P_H$ for different configurations of the device:
$R_{QPC1,2,3}=$ \SI{7.4}{k\ohm} (b), $R_{QPC1,2}=$ \SI{20}{M\ohm}, $R_{QPC3}=$ \SI{1}{M\ohm} (c).
For each case, the full lines are a fit of the expressions given in the text.
(d) Joint bi-logarithmic plot of (b) and (c).}
\end{figure} 

After demonstrating the functionality of this thermometry scheme, we turn to the investigation of heat relaxation mechanisms in the microdomain. At steady state, the temperature $T_2$ is the result of the following thermal-balance equation: 
\begin{equation}\label{powerranger}
\sum_{i=1}^3{\dot{Q}_{QPCi}}+\dot{Q}_{QD}+\dot{Q}_{e-ph}=0\ ,
\end{equation}
where $\dot{Q}_{e-ph}$ is the power exchanged between electrons and the phonon bath assumed to be at equilibrium at the cryostat base temperature ($T_{ph}=T_{bath}=\SI{250}{mK}$),
while $\dot{Q}_{QPCi}$ and $\dot{Q}_{QD}$ are the heat currents flowing through the i-th QPC and the QD, respectively
\footnote{Power flowing out of the domain is given a positive sign.}. A schematic drawing of these contributions is shown in Fig.~4(a). We now briefly discuss each term.

The power flow $\dot{Q}_{QPCi}$ through each QPC can be calculated within the Landauer-B\"uttiker formalism \cite{Sivan1986,VanHouten1992}.
If the transmission coefficient is energy-independent in the relevant range (which also results in a linear current-voltage characteristic), the result can be evaluated analytically. For the biased QPCs it takes the form
$\dot{Q}_{QPC1}=\dot{Q}_{QPC2}=-\frac{V_HI_H}{4}+\frac{L_0}{2 R_{QPC1,2}}\left[T_2^2-T_1^2\right]$,
where $V_H$ is the total voltage drop developed across the heating circuit, $L_0=\pi^2k_B^2/3e^2$ is the Lorenz number and we made use of the fact that $R_{QPC1}=R_{QPC2}$.
The first term in this expression
is reminiscent of Joule heating: of the total power $V_H I_H$ provided by the current source, one half is delivered into the domain, $P_H=I_HV_H/2$. The second term is related to heat leakage from the domain to the (colder) adjacent 2DEG region through quasiparticle tunneling. It is proportional to the difference of the squared temperatures, in agreement with the Wiedemann-Franz law \cite{Switkes1998}. For the unbiased QPC, only the latter term contributes so that $\dot{Q}_{QPC3}=(L_0/2 R_{QPC3})\left[T_2^2-T_1^2\right]$.

Let us now turn to $\dot{Q}_{e-ph}$, the power transferred through electron-phonon coupling. At low temperatures, heat exchange between the 2DEG and the three-dimensional phonon bath mainly relies on the scattering with low-energy acoustic phonons. In particular, for III-V semiconductor alloys, piezoelectric coupling is expected to dominate over deformation-potential interaction \cite{Price1982} at subkelvin temperatures. The energy-loss rate for the screened piezoelectric electron-phonon interaction in the clean limit is given by the expression
$\dot{Q}_{e-ph}= \Sigma A_D \left( T_2^5-T_{ph}^5 \right)$, where
the coupling constant $\Sigma$ depends on the electron density and on host-crystal material parameters
\footnote{Namely, $\Sigma=n_s [(\zeta(5)m^{*2}e^2h_{14}^2)/(64\,\pi(2\pi\, n_s)^{3/2}\hbar^7\rho\, q_s^2)]$ $\times\left[(135/s_l^4)+(177/s_t^4)\right]k_B^5$, where $\rho$ is the  mass density, $h_{14}$ the piezoelectric coupling constant, $s_l$ ($s_t$) the longitudinal (transverse) sound velocity, $q_s=m^*e^2/2\pi\hbar^2\epsilon_r\epsilon_0$ is the Thomas-Fermi screening wavevector, and $\epsilon_r$ and $\epsilon_0$ are the relative and vacuum permittivity, respectively.}.
For our GaAs/AlGaAs 2DEG we obtain $\Sigma \approx \SI{30}{fW\micro m^{-2} K^{-5}}$
\footnote{We used the elastic parameters reported in \cite{Ma1991} and included a 0.77 correction factor to account for the phonon spectrum anisotropy in GaAs \cite{Jasiukiewicz1996}.}.

In light of the foregoing discussion, the temperature data of Fig.~3(c) are re-plotted in Fig.~4(b) as a function of the injected power $P_H$, evaluated from the measured $V_H$ and $I_H$.
The data are well described by the power law $P_H=A+B(T_2^2-T_1^2)$, where $T_1=$ 400 mK and the constant $A$ accounts for spurious heating. The best fit of this expression yields $A=(-0.15\pm 0.02)\SI{}{pW}$ and $B=(5.0\pm 0.1)\SI{}{pW K^{-2}}$. 
Notably, the value for $B$ agrees with the \SI{4.9}{pW K^{-2}} predicted for the heat leakage through three QPCs of equal resistance \SI{7.4}{k\ohm}. In this regime, heat flow through the QPCs dominates over electron-phonon heat exchange.
By increasing the resistance of the QPCs, one can make the former contribution negligible. The data plotted in Fig.~4(c) were taken for a device configuration in which the QPC1 and QPC2 zero-bias resistances were set to \SI{20}{M\ohm}, while that of QPC3 to \SI{1}{M\ohm}. The continuous line is a fit to the data of the power law $P_H=C+D(T_2^5-T_{ph}^5)$. 
The good agreement with the experimental data demonstrates the transition to a regime where heat relaxation relies on coupling to the phonon bath. This is manifest in the joint log-log plot of Fig.~4(d), where the two datasets are characterized by different exponents as well as by a different order of magnitude for $P_H$. From the least-squares fit we extract $C=(-5\pm 1)\SI{}{fW}$ and $D=(170\pm 20)\SI{}{fW K^{-5}}$.  Since $A_D=\SI{16}{\micro m^2}$, this yields $\Sigma=(11\pm 1)\SI{}{fW \micro m^{-2} K^{-5}}$. The obtained value is about three times smaller than the theoretical prediction and about two times smaller than what was previously measured for a much larger electron reservoir \footnote{The electron-acoustic phonon coupling constant of GaAs/AlGaAs in the subkelvin regime and in the absence of magnetic field was first measured by Appleyard \textit{et.~al.} \cite{Appleyard1998}. We compare 2DEGs of different densities according to $\Sigma \propto n_s^{-1/2}$.}.
The reasons for such discrepancy are not completely clear. For our case, possible sources of uncertainty are related to 
the estimate of the effective area relevant for the electron-phonon interaction and of the average electron density inside the reservoir.

In conclusion, we have demonstrated local electronic thermometry based on a lateral QD, shown that heat exchange between the 2DEG and phonons depends on temperature according to a power law consistent with the screened electron-acoustic phonon piezoelectric interaction, and provided a measurement of the electron-phonon coupling constant in the subkelvin regime.
Such results are relevant to the implementation of nanodevices which rely on the manipulation of thermal distributions, e.g. in electronic refrigeration schemes based on QDs \cite{Prance2009,Edwards1993,Edwards1995}. The method here reported can be readily applied to other material systems characterized by different electronic and phononic properties, e.g., InGaAs alloys with high In content \cite{Deon2010} or semiconductor nanowires \cite{DeFranceschi2003}.

We gratefully acknowledge F. Dolcini and S. De Franceschi for fruitful discussions.
The work was partially supported by the INFM-CNR Seed project `Quantum Dot Refrigeration: Accessing the \SI{}{\micro K} Regime in Solid-State Nanosystems', by the NanoSciERA project `NanoFridge', and by the European Community FP7 project No. 228464 `Microkelvin'.


%

\end{document}